\documentclass[iop,apj,tighten]{emulateapj}
\usepackage{amsmath,amstext}
\usepackage[breaklinks,colorlinks,citecolor=blue,linkcolor=magenta]{hyperref} 
\usepackage[all]{hypcap} 
\newcommand{\msun}{\ensuremath {\rm M}_{\odot}}

\newcommand{\kms}{\ensuremath{ {\rm km s}^{-1}}}
\shorttitle{Dynamics of BH Binaries}
\shortauthors{O'Leary, Meiron, \& Kocsis}

\begin{document}
\title{Dynamical formation signatures of black hole binaries in the first detected mergers by LIGO}

\author{Ryan M.~O'Leary\altaffilmark{1}, Yohai Meiron\altaffilmark{2}, and Bence Kocsis\altaffilmark{2}}
\affil{$^1$JILA, University of Colorado and NIST, 440 UCB, Boulder, 80309-0440, USA; \url{ryan.oleary@jila.colorado.edu}}
\affil{$^2$Institute of Physics, E\"otv\"os University, P\'azm\'any P. s. 1/A, Budapest, 1117, Hungary}

\begin{abstract}
The dynamical formation of stellar-mass black hole-black hole binaries has long been a promising source of gravitational waves for the Laser Interferometer Gravitational-Wave Observatory (LIGO). 
Mass segregation, gravitational focusing, and multibody dynamical interactions 
naturally increase the interaction rate between the most massive black holes in dense stellar systems, eventually leading them to merge.  We find that dynamical interactions, particularly three-body binary formation, enhance the merger rate of black hole binaries with total mass $M_{\rm tot}$ roughly as $\propto M_{\rm tot}^{\beta}$, 
with $\beta \gtrsim 4$. We find that this relation holds mostly independently of the initial mass function, but the exact value depends on the degree of mass segregation.
The detection rate of such massive black hole binaries is only further enhanced by LIGO's greater sensitivity to massive black hole binaries with $M_{\rm tot}\lesssim 80\,\msun$. 
We find that for power-law BH mass functions $\mathrm{d}N/\mathrm{d}M \propto M^{-\alpha}$ with $\alpha \leq 2$, LIGO is most likely to detect black hole binaries  with a mass twice that of the maximum initial black hole mass and a mass ratio near one.
Repeated mergers of black holes inside the cluster result in about $\sim 5\,\%$ of mergers being observed between two and three times the maximum initial black hole mass. 
Using these relations, one may be able to invert the observed distribution to the initial mass function with multiple detections of merging black hole binaries.
\end{abstract}

\maketitle

\section{Introduction}

After over two decades of development, the  Advanced Laser Interferometer Gravitational-Wave Observatory\footnote{\url{http://www.ligo.org/}} (aLIGO) has directly detected gravitational waves from an inspiralling black hole-black hole (BH-BH) binary \citep{2016PhRvL.116f1102A}. 
Through the precise measurement of the gravitational waves, aLIGO is capable of characterizing many properties of inspiralling binaries, including the total mass of the binary, $M_{\rm tot}$, the mass ratio, $q$, and the black holes' spins. The first detected BH-BH binary, GW150914, had unusually high component masses  $(m_1,m_2)=(36^{+5}_{-4}\msun,29^{+4}_{-4}\msun)$ in comparison to BH masses inferred for star-BH X-ray binaries \citep{Farr11,2015ApJ...800...17F}. A second, less significant event\footnote{LVT151012 has a false alarm probability of 0.02.}, LVT151012, also had high inferred masses $(m_1,m_2)=(23^{+18}_{-5}\msun,13^{+4}_{-5}\msun)$ \citep{LVTpaper}. aLIGO has finally opened a new window to our Universe. Along with other upcoming instruments VIRGO\footnote{\url{http://www.virgo-gw.eu/}} and KAGRA\footnote{\url{http://gwcenter.icrr.u-tokyo.ac.jp/en/}}, aLIGO will allow us to probe the demographics of potentially hundreds of BH-BH binaries \citep{LIGOApJL}.

There are three primary pathways that lead to BH-BH binaries that can merge within the age of the Universe, through binary evolution, through gas dynamics \citep{2011ApJ...740L..42D,2016arXiv160203831B,2016arXiv160204226S}, and through stellar dynamics \citep[see][for a review]{2013LRR....16....4B}.
First, such binaries can form through the evolution of isolated, massive binary stars.\footnote{or perhaps the core collapse of a single supermassive star \citep{2013PhRvL.111o1101R,2016ApJ...819L..21L,2016arXiv160300511W}}  
A major bottleneck in our understanding of this channel is the complex tidal \citep{Mandel16,Marchant16,2016arXiv160302291D} and common envelope \citep[e.g][]{Belczynski15} evolution such a binary must go through in order to produce two BHs that can merge within a Hubble time from the loss of gravitational waves. This is in addition to uncertainties in the details of massive star evolution, supernova explosions, and the birth kicks of black holes.  Nevertheless, sophisticated population synthesis routines have been developed that incorporate many of these uncertainties to make predictions about the properties of the first gravitational wave sources \citep[e.g.,][]{Belczynski2008ApJS..174..223B,Belczynski2010ApJ...715L.138B,2012ApJ...759...52D,2014A&A...564A.134M}. A second possibility to get BH binary mergers is through gas assisted mergers \citep{2005ApJ...630..152E,2011ApJ...740L..42D}, however simple rate estimates suggest that gas assisted stellar BH mergers are probably relatively uncommon \citep{2016arXiv160203831B,2016arXiv160204226S}. 

Dynamical interactions of BHs in dense stellar environments, such as globular clusters, present another promising method to produce tight BH-BH binaries whether through exchange interactions \citep{PZM00}, three body dynamics \citep{millerhamilton02,2003ApJ...598..419W,Antonini2014ApJ...781...45A,Antognini2014MNRAS.439.1079A}, or direct dynamical capture\footnote{see \citet{2016arXiv160300464B} for GW captures of BHs that constitute dark matter} \citep{1989ApJ...343..725Q,1993ApJ...418..147L,OKL09}.  In these scenarios, the black holes that form at the end stage of stellar evolution collect near the center of the cluster through dynamical friction \citep{1993Natur.364..421K,1993Natur.364..423S}.   Because of gravitational focusing, the most massive BHs are preferentially involved in subsequent gravitational encounters and form BH-BH binaries. These binaries may merge within the cluster \citep{millerhamilton02b} or be ejected from the cluster and merge on much longer timescales \citep{OLeary06}. Such models have presented their own theoretical obstacles, the initial mass function of BHs perhaps the largest, but ever sophisticated simulations over nearly a decade  have generally found similar estimates for the expected merger rate of the binaries as well as their characteristics  \citep{2006ApJ...640..156G,moodysigurdsson,2010MNRAS.402..371B,2010MNRAS.407.1946D,2011MNRAS.416..133D,2013ApJ...763L..15M,2014MNRAS.440.2714B,2014MNRAS.441.3703Z,2015ApJ...800....9M,2015PhRvL.115e1101R}.   These results remain even in simulations that have shown a substantial fraction of BHs remain in the cluster \cite[e.g.][]{Mackey2008MNRAS.386...65M,2015ApJ...800....9M,2015PhRvL.115e1101R}

In this work, we present a number of observational signatures of the dynamical formation of black hole binaries.  In particular, we focus on signatures that are independent of the poorly known black hole initial mass function (IMF) \citep{2004ApJ...611.1068B,Ozel10,Farr11,Belczynski15}. 
A number of studies have qualitatively discussed that dynamical interactions preferentially form binaries with the most massive components in the cluster \citep[e.g.][]{OLeary06,OLeary07,2015ApJ...800....9M}. Yet few studies focused on the expected mass ratio distribution of the BH-BH binaries that merge. In this work, we use the original Monte Carlo results of \citet{OLeary06}, a series of new Monte Carlo simulations, as well as a series of new direct $N$-body simulations to explore the expected mass distribution of the components of BH-BH binaries.   We argue that the mass distribution of the BH binaries found by aLIGO will present a unique signature of dynamically formed binaries and their underlying mass function.  
After we have submitted our manuscript two papers appeared on the expected rates of stellar black hole mergers in globular clusters with independent methodologies, which confirm our findings \citep{2016arXiv160202444R,2016arXiv160300884C}

\section{Methods}
After the first supernovae, the more massive BHs collect near the center of the cluster owing to dynamical friction from the low mass stellar background. In contrast to previous expectations \citep{1993Natur.364..421K,1993Natur.364..423S}, however, these BHs do not interact exclusively amongst themselves.  After the BH subcluster experiences core collapse and produces a binary, that binary acts through the BH population to heat the entire star cluster until it is ejected from the system (\citealt{Mackey2008MNRAS.386...65M}; cf. \citealt{osm14}). Nevertheless, the strongest dynamical interactions that eventually lead to binary ejection rarely include stars \citep{2015ApJ...800....9M}.  In this work, we are interested in the properties of the BH-BH binaries, and only follow the evolution of the BHs separately from the cluster.
Although this does not follow the proper evolution of the entire BH cluster, it does capture the important dynamical interactions that lead to binary formation and, eventually, binary heating.
In this work, we use two methods for following the evolution of the BHs in dense stellar clusters. 
We use the Monte Carlo Method presented in \citet{OLeary06} as well as a direct $N$-body method \citep{Wang15} that follows the evolution of only the BHs. Both methods focus on the evolution of the BHs in isolation from the stellar cluster. 

Our Monte Carlo simulations of BHs in globular clusters follow directly from \citet{OLeary06}, and is based on the original method by \citet{2005MNRAS.358..572I,2010ApJ...717..948I}. Globular clusters are treated as two component systems, with a constant density core, where all dynamical interactions take place, and a low-density halo.  BHs that are kicked into the halo from dynamical interactions remain there until dynamical friction returns the BHs to the cluster center.  The three and four-body interactions  in the cluster are sampled from their likelihood distributions and directly integrated with {\sc fewbody} \citep{2004MNRAS.352....1F}.  Binaries that are present in the simulation are evolved following \citet{1964PhRv..136.1224P} equations which describes the inspiral due to gravitational wave (GW) emission.  
If two non-spinning BHs merge within the cluster their merger product receives a kick due to the asymmetric emission of GWs 
with a maximum velocity of $175\,\kms$ near $q \equiv M_1/M_2 \approx 1/5$ \citep{Gonzalez07}. Equal mass mergers receive no kick as they are symmetric.  
The BH mass function changes in the simulation due to BH mergers and ejections.

We also run a series of direct $N$-body simulations to follow the evolution of the BHs that form in a much larger stellar cluster. We integrate the orbits of all the BHs directly using {\sc nbody6++} \citep{Wang15}, and keep track of the dynamically-formed binary population. Future work will include the full post-Newtonian treatment throughout the cluster evolution, as mergers in the cluster likely constitute about $\sim 15\,\%$ of all mergers \citep{2015ApJ...800....9M}. These simulations, due to their initial conditions, produce too few BH-BH binaries that merge in a Hubble time, so we include all binaries ejected from the system in our analysis to compare with the Monte Carlo results. Two model suites were produced. First, the model that includes a potential that mimics the underlying stellar cluster using a static Plummer sphere with half-mass radius $R_{\rm h} = 13\,$pc and mass $M_{\rm cl} = 2\times 10^{5}\,\msun$ and a BH subcluster made of 492 single particles and 10 ``primordial'' binary pairs with half-mass radius of 3.3\,pc.
We also simulate the cluster of BHs without an external potential. This model has only 512 single particles taken from a Plummer distribution with half-mass radius of 0.5\,pc. The integration is performed to 10\,Gyr in all models.

\begin{figure*}
\includegraphics[width=0.49\textwidth]{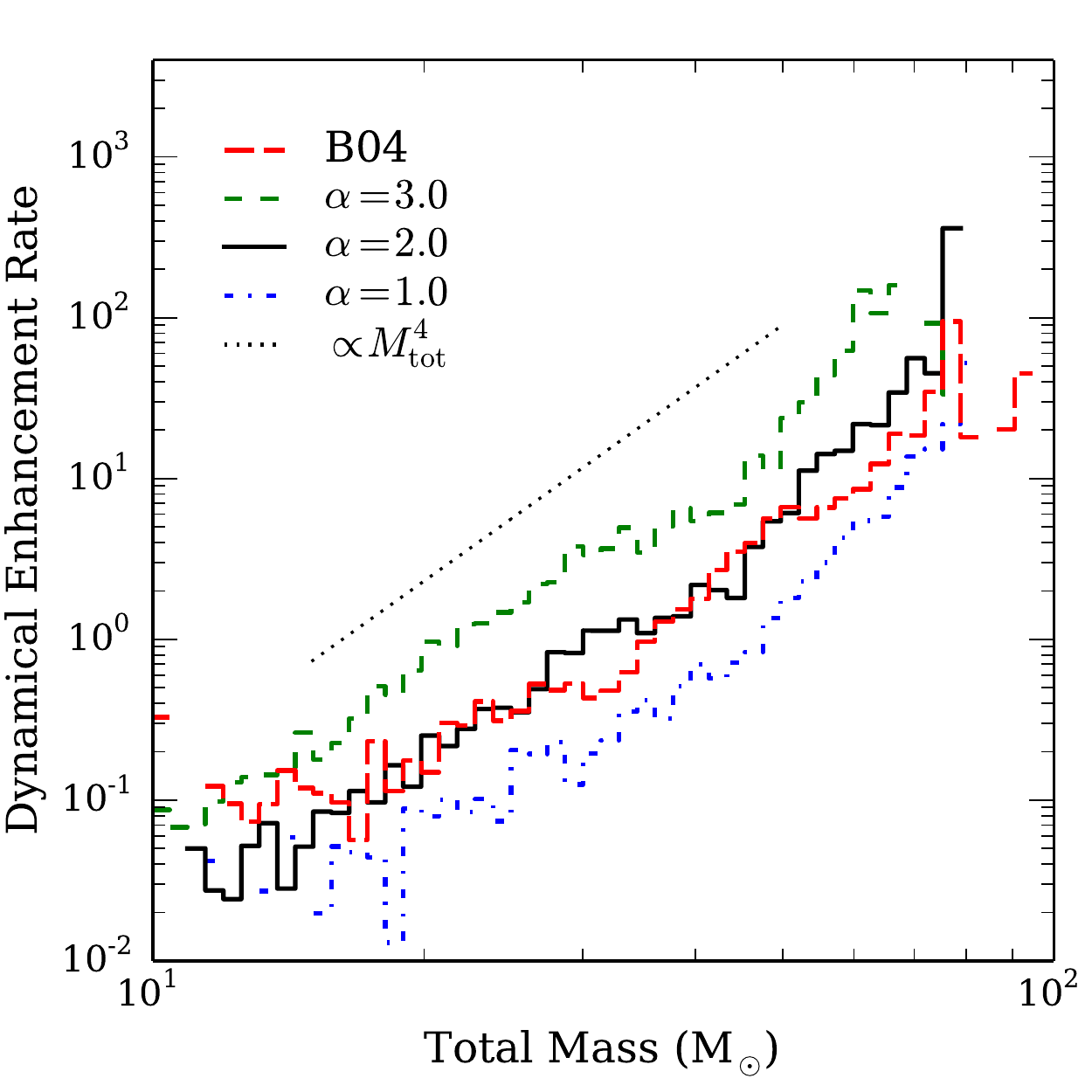}
\includegraphics[width=0.49\textwidth]{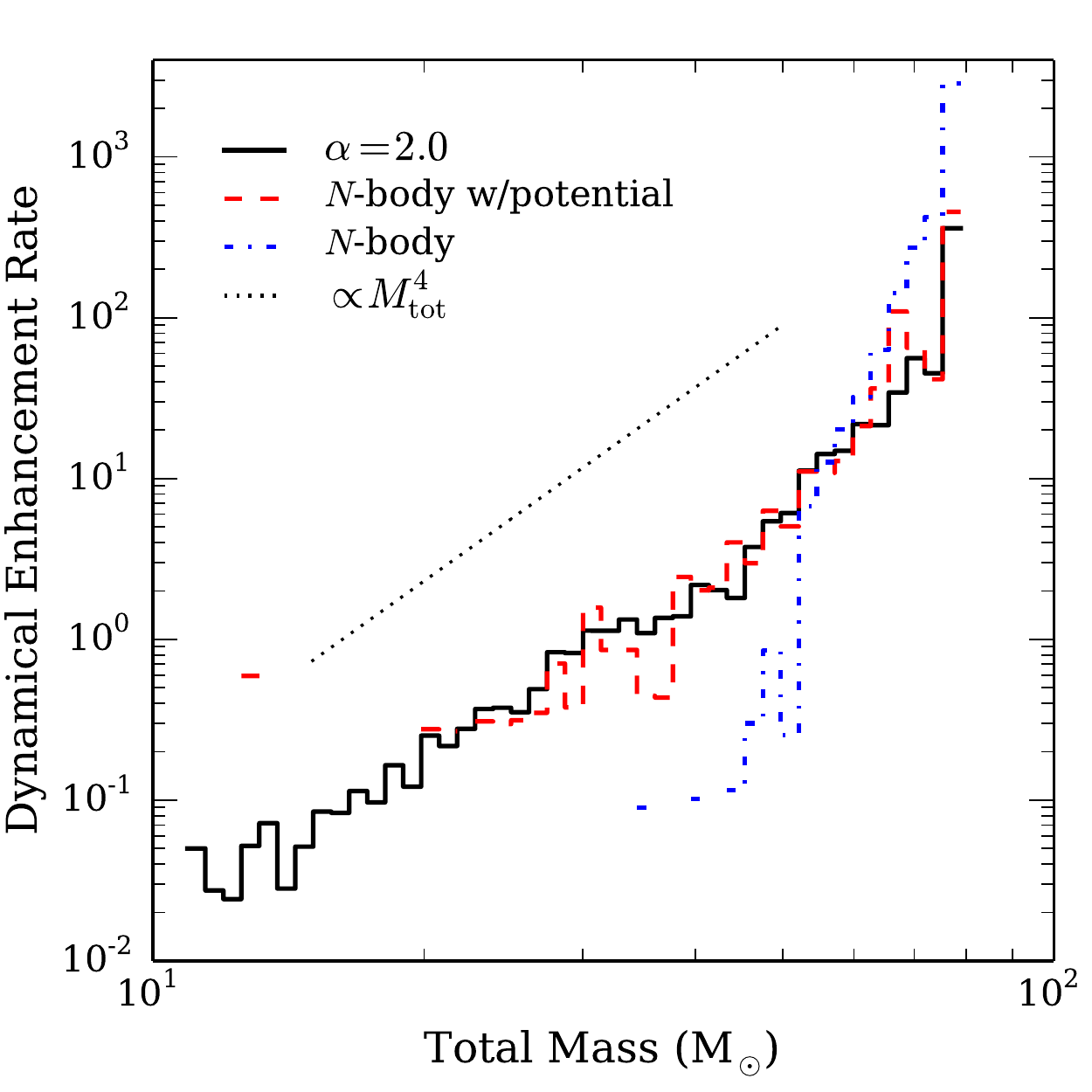}

\caption{\label{fig:massto} Dynamical effects on the total binary mass distribution of BH-BH binaries. 
We plot the ratio between the mass distribution of BH-BH binaries that merge in a Hubble time, and the mass distribution of BH-BH binaries if they randomly sampled the initial BH mass function (Eq.~\ref{eq:mf}.)  
{\em Left}: We show this ratio for four mass functions with our Monte Carlo simulations.  Three of the mass functions are power-law mass functions $M^{-\alpha}$ with slope $\alpha = 1$, 2, and 3 (blue dot-dashed, black solid, and green dashed lines, respectively), and $M_{\rm max} = 40\,\msun$.  Also shown is the ratio for the \citet{2004ApJ...611.1068B} mass function for BHs that formed from low-metallicity progenitors (red long-dashed line).  For all of the mass functions, we find that ratio scales roughly as $M_{\rm tot}^{\beta}$ with $\beta \approx 4$ (black dotted line).  Near $M_{\rm max}$, the dynamically formed BH-BH mergers are approximately $\gtrsim 50$ times larger than expected from the mass distribution alone. Most black holes in this diagram encountered at most one merger, and the mergers of BHs which have grown beyond $M_{\rm max}$ are not shown.
{\em Right}: We show the ratio of BH-BH total masses to a random distribution from our $N$-body simulations. The two $N$-body simulations use the same mass function  as our Monte Carlo simulation (solid black line; $M_{\max} = 40\,\msun$ and $\alpha = 2$).  The red dashed line shows the binary distribution from our $N$-body simulation with the external potential. The blue dash-dotted line show the results from our simulation without the external potential. Far fewer binaries form with $M_{\rm tot} < M_{\rm max}$.}
\end{figure*}

In both Monte Carlo and $N$-body simulations, we generate our initial BH mass function from a power-law distribution
\begin{equation}
\label{eq:mf}
f(M)~\equiv~\frac{\mathrm{d}N}{\mathrm{d}M}~\propto M^{-\alpha},
\end{equation}
between minimum and maximum BH mass of $M_{\min}=5\,\msun$ and $M_{\max}=40\,\msun$
with $\alpha\,=\,$0, 1, 2, 3, and 4. While this range of masses is larger than the inferred masses of BHs in {\em X}-ray binaries \citep{Ozel10,Farr11}, it is consistent with models of BHs that form with fallback accretion and direct collapse \citep{2004ApJ...611.1068B,Belczynskimaxmass2010ApJ...714.1217B}, and is minimally required to explain GW150914. 
We also run a set of simulations using the BH mass function found in the population synthesis studies of \citet{2004ApJ...611.1068B} for low-metallicity stars. This mass function has three peaks in its distribution, near $7\,\msun$, $14\,\msun$, and $24\,\msun$, and represents a more complicated and perhaps more realistic distribution of masses.  In the Monte Carlo simulations, we assume that $10\,\%$ of black holes are in primordial binaries (cf. the initial conditions for the $N$-body above), with a period distribution that is flat in the log of period.

These two methods are complimentary to each other, and both capture the strongest dynamical encounters between BHs.  Nevertheless they have their limitations and benefits.  In both series of simulations, we do not directly follow the entire coupling between the main stellar population in the cluster and the black holes, which is important to determine the present day population of BHs in globular clusters \citep[e.g.,][]{Mackey2008MNRAS.386...65M,2015ApJ...800....9M,Rodriguez16,Wang16}.
However, the latter simulations confirm that the dynamical interactions that lead to binary ejections are driven exclusively by the BH subcluster, and are dominated by BH-BH binaries.  With the fast Monte Carlo simulations, we can run a much larger number of simulations with different compositions and initial conditions than studies which follow the evolution of the entire stellar population, and we can turn important processes, such as three-body binary formation, off in order to understand which aspects of the dynamical interactions are most important.

To leading (2.5 post-Newtonian) order, the root-mean-square (RMS) detection signal to noise ratio for a circular inspiraling binary is given as
\begin{equation}\label{eq:SNR}
\frac{S}{N} = k
\frac{\eta^{1/2}  M_{\rm tot}^{5/6}}{d_{\rm L}}I_7^{1/2}(M_{\rm tot})
\end{equation}
where $k=\pi^{-2/3}\sqrt{2/15} \approx 0.17$ for an isotropic binary orientation relative to the detector, $\eta=q/(1+q)^2$ is the symmetric mass ratio, $d_{\rm L}$ is the luminosity distance, and
\begin{equation}
I_7(M) = \int^{f_{\rm ISCO}(M)}_{f_{\rm min}}\frac{f^{-7/3}}{S_h(f)}\,df
\end{equation}
where $f_{\rm ISCO}(M_{\rm tot})=\pi^{-1}6^{-3/2}M^{-1}$ is the GW frequency at the innermost stable circular orbit, $f_{\min}$ is the lowest GW frequency that the instrument can detect, and $S_h$ is the one-sided noise spectral density \citep{1994PhRvD..49.2658C,2006PhRvD..74f3006D}.\footnote{We adopt units ${\rm G}={\rm c}=1$. To convert from mass to time units, one should multiply by ${\rm G}/{\rm c}^3$. If the source is at a cosmological distance, redshift $z$, then $M_{\rm tot}$ should be replaced by $(1+z)M_{\rm tot}$.} We set $f_{\rm min}=10\,$Hz and $S_h(f)$ using the calibrated sensitivity spectra of aLIGO-Hanford on October 1, 2015 \citep{LIGOO1sensitivity}. For a fixed detection threshold (e.g. $S/N\geq 7$) the maximum distance range $d_{\rm L,max}$ may be obtained from Eq.~(\ref{eq:SNR}). 
Assuming that the source population is uniformly distributed in volume, the detection rate is biased by approximately
\begin{equation}\label{eq:V}
V_{\rm det} \propto d_{\rm L,max}^3 \propto \frac{q^{3/2}}{(1+q)^{3}}M_{\rm tot}^{5/2}I_7^{3/2}(M_{\rm tot})\,.
\end{equation}
This function grows as $M_{\rm tot}^{2.3}$ between $M\sim 10$--$20\,\msun$, has a maximum at $M_{\rm tot}=77\,\msun$ and decreases to zero at $439\,\msun$ where $f_{\rm ISCO}=f_{\rm min}=10\,$Hz. Compared to $10\,\msun$ and fixed $q$, the detectable volume is a factor 27 larger for $M_{\rm tot}=77\,\msun$ and a factor 4.5 smaller at $400\,\msun$.

\begin{figure*}
\includegraphics[width=0.49\textwidth]{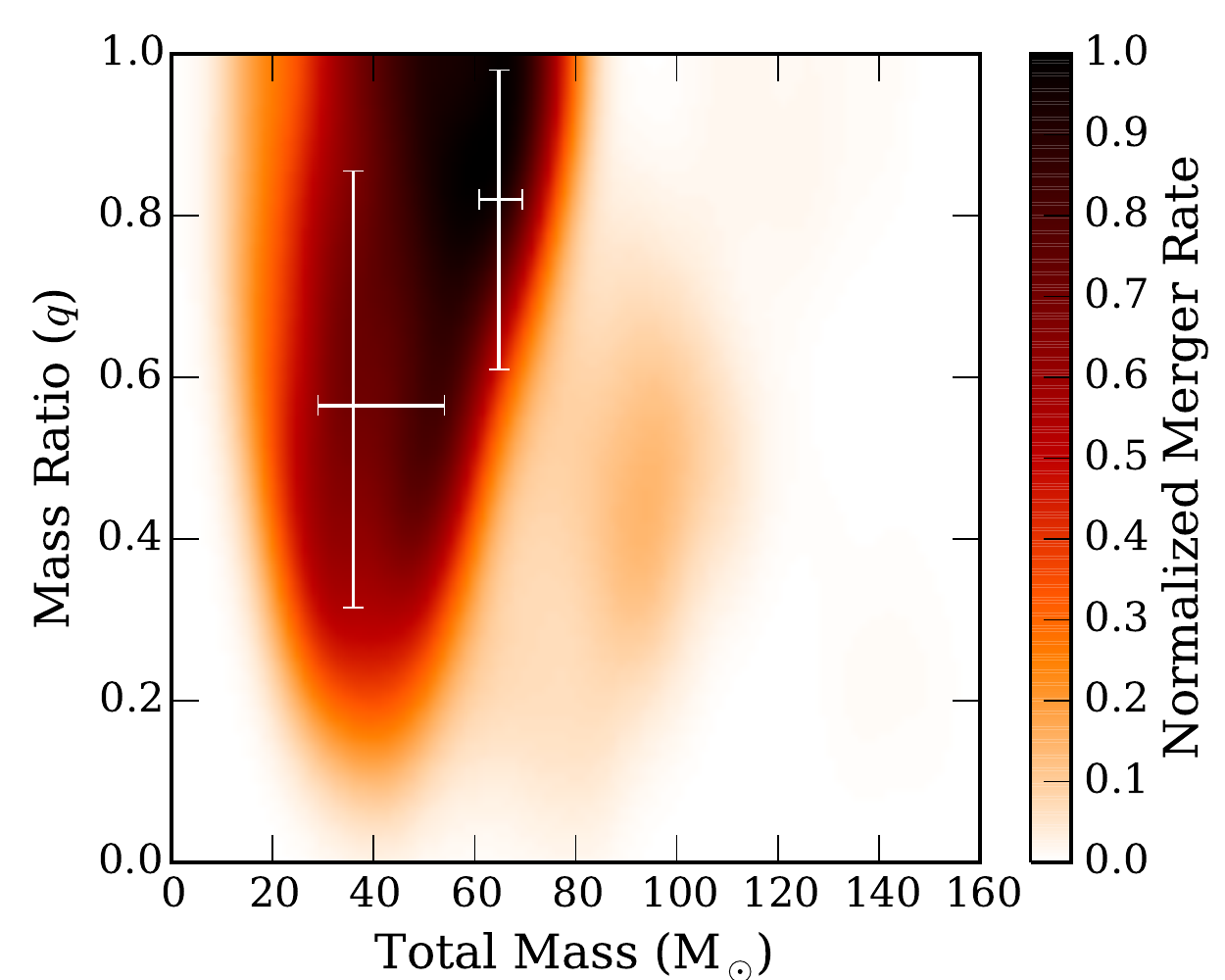}
\includegraphics[width=0.49\textwidth]{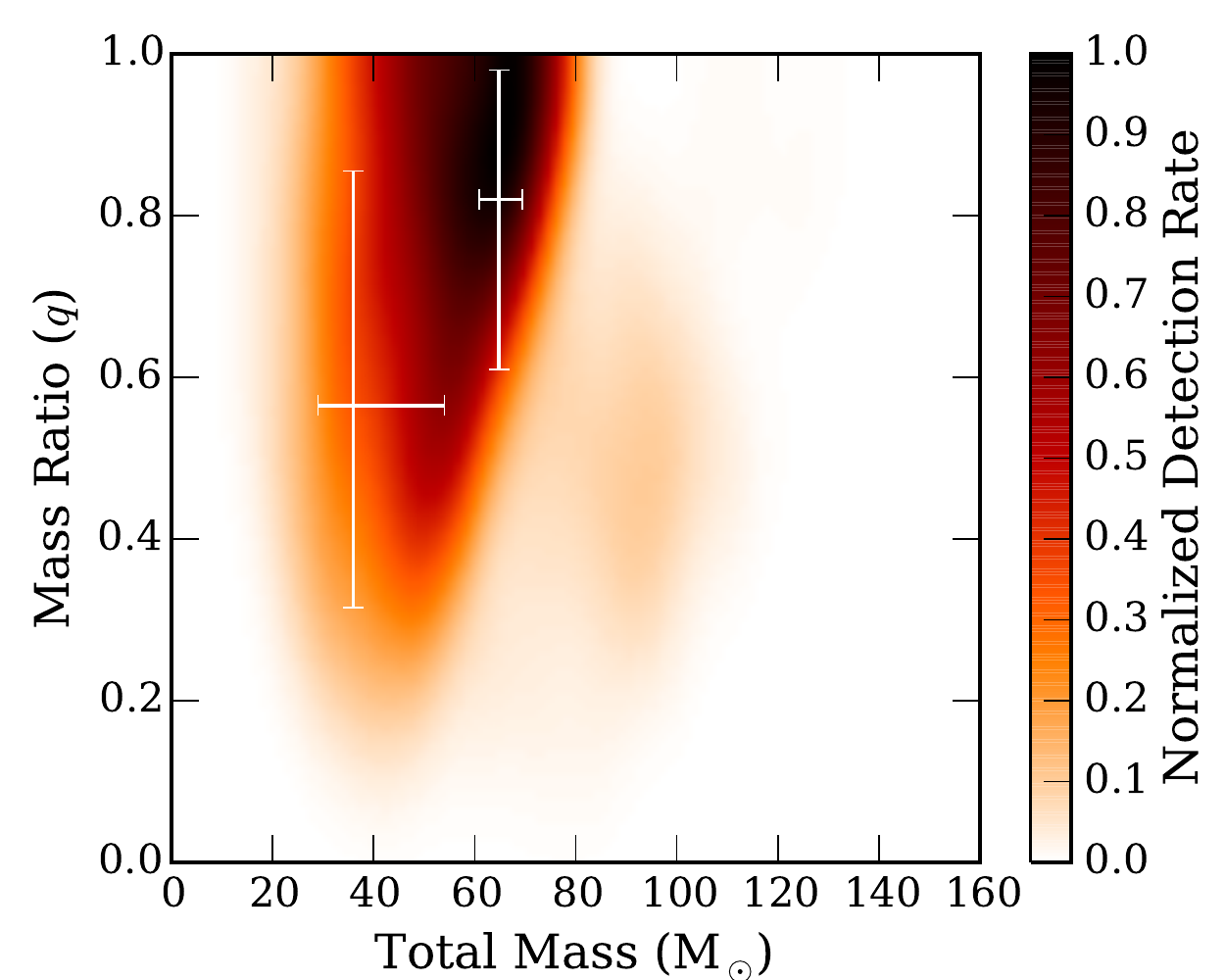}\\
\includegraphics[width=0.49\textwidth]{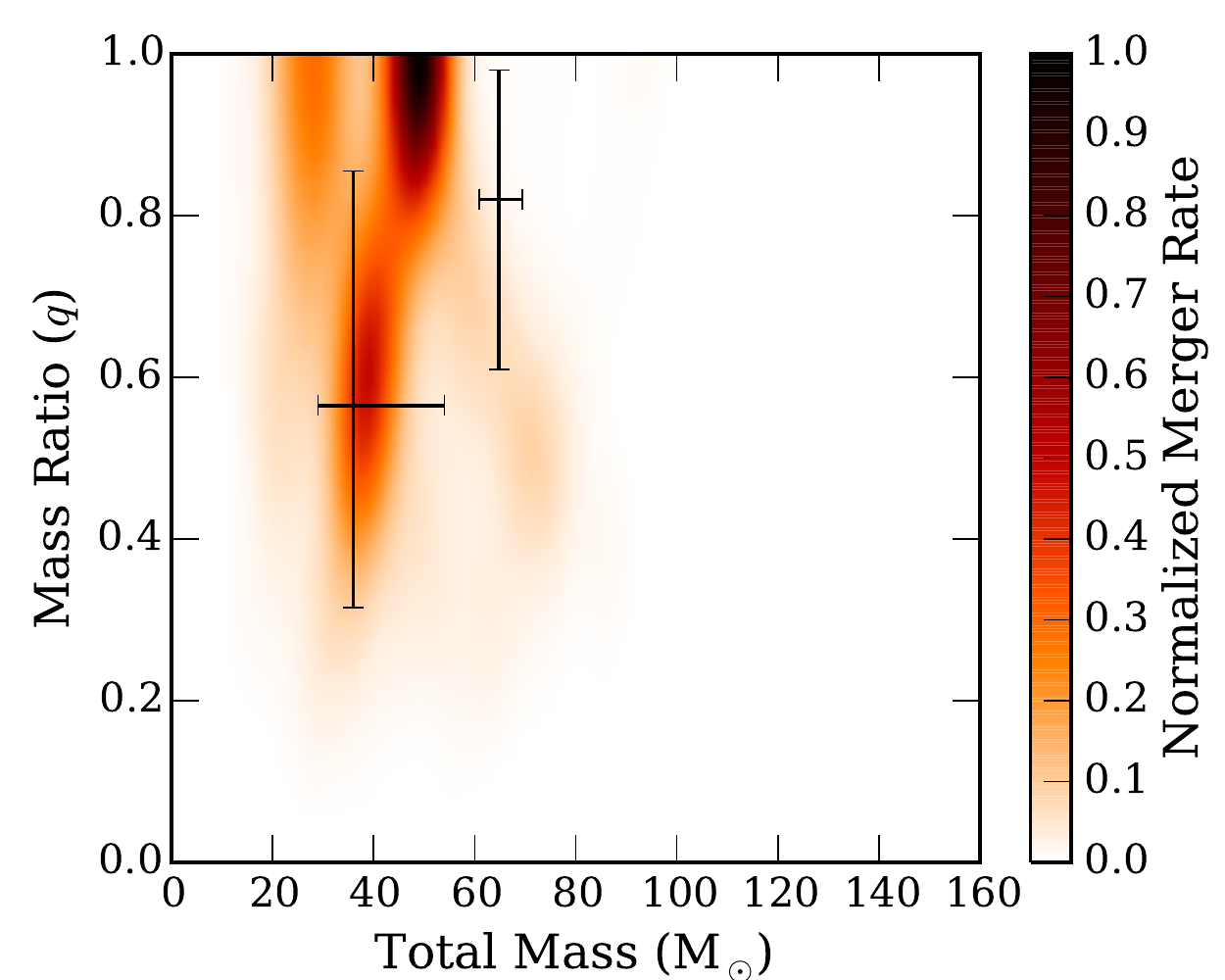}
\includegraphics[width=0.49\textwidth]{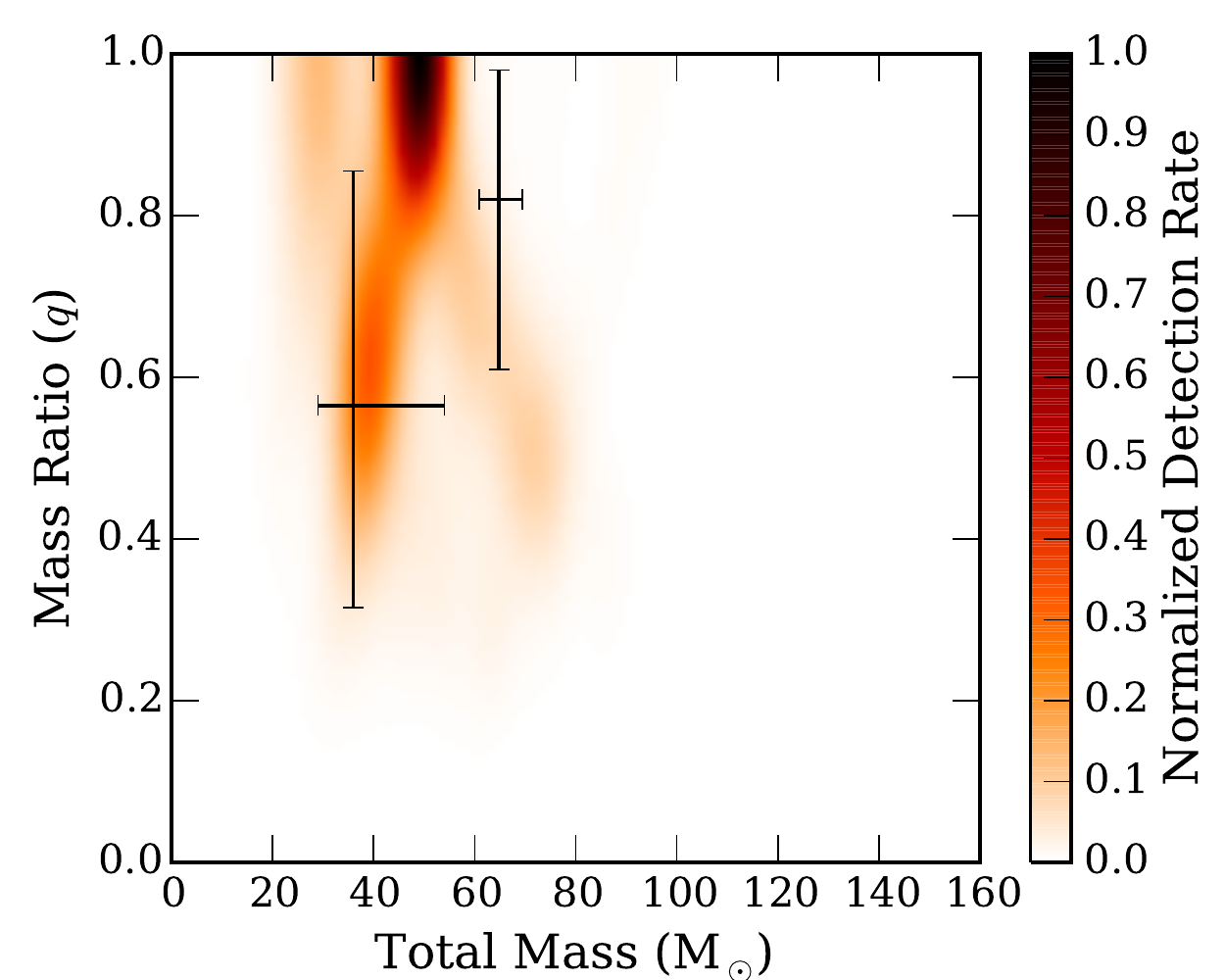}\\
\caption{\label{fig:massrat} Total binary mass and mass ratio distributions for inspirals for an $\alpha=2.0$ power-law mass function ({\em top panels}) and the \citet{2004ApJ...611.1068B} mass function ({\em bottom panels}). {\em Left panels:} The distribution of dynamically formed BH-BH binaries that merge within $10\,$Gyr. 
{\em Right:} Normalized detection rate of inspiralling BH-BH binaries for all star clusters within the detectable Universe for the O1 run of LIGO Hanford. The errorbars show the $90\%$ confidence interval for the properties of GW150914 and LVT151012.}
\end{figure*}

\section{Results}
Dynamical interactions between BHs in stellar clusters are primarily driven by mass segregation,  gravitational focusing, and multibody dynamics.
We therefore expect the most massive BHs to be preferentially found in merging BH-BH binaries. 

In Figure~\ref{fig:massto}, we compare the mass distribution of merged BH-BH binaries from our simulations to that expected from random pairings of BHs from the BH mass function.  We find that dynamical interactions enhance the merger rate of BH binaries relative to random pairings in a star cluster by a boost factor that scales as a power-law,
\begin{equation}
\label{eq:enhance}
B = k_1 M_{\rm tot}^\beta
\end{equation}
with $\beta \gtrsim 4$ and we assume that $k_1$ is independent of $(q,M_{\rm tot})$. All of our Monte Carlo simulations have $\beta \approx 4$. This is even true for the \citet{2004ApJ...611.1068B} mass function, which is not monotonically decreasing as a function of mass, and has multiple peaks.    We have also found similar results for our Monte Carlo simulations with a variety of cluster models, with differing velocity dispersions and escape velocities.   Although our $N$-body simulations with an external potential still had $\beta \approx 4$, we found that the binaries that form in the cluster without the external potential have $\beta \gg 4$.  
 We discuss this in more detail in Section~\ref{sec:conc}.

\begin{figure*}
\includegraphics[width=0.49\textwidth]{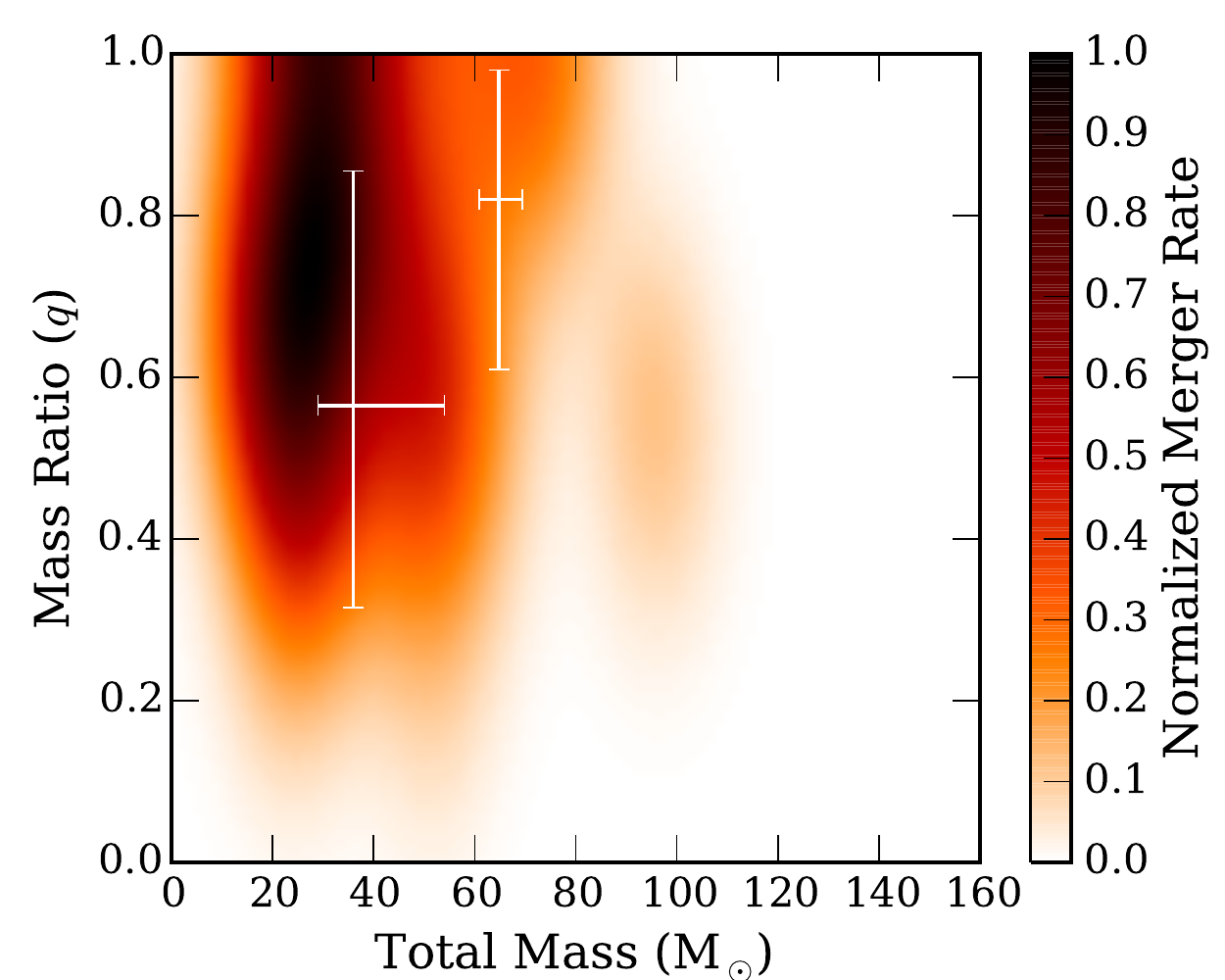}
\includegraphics[width=0.49\textwidth]{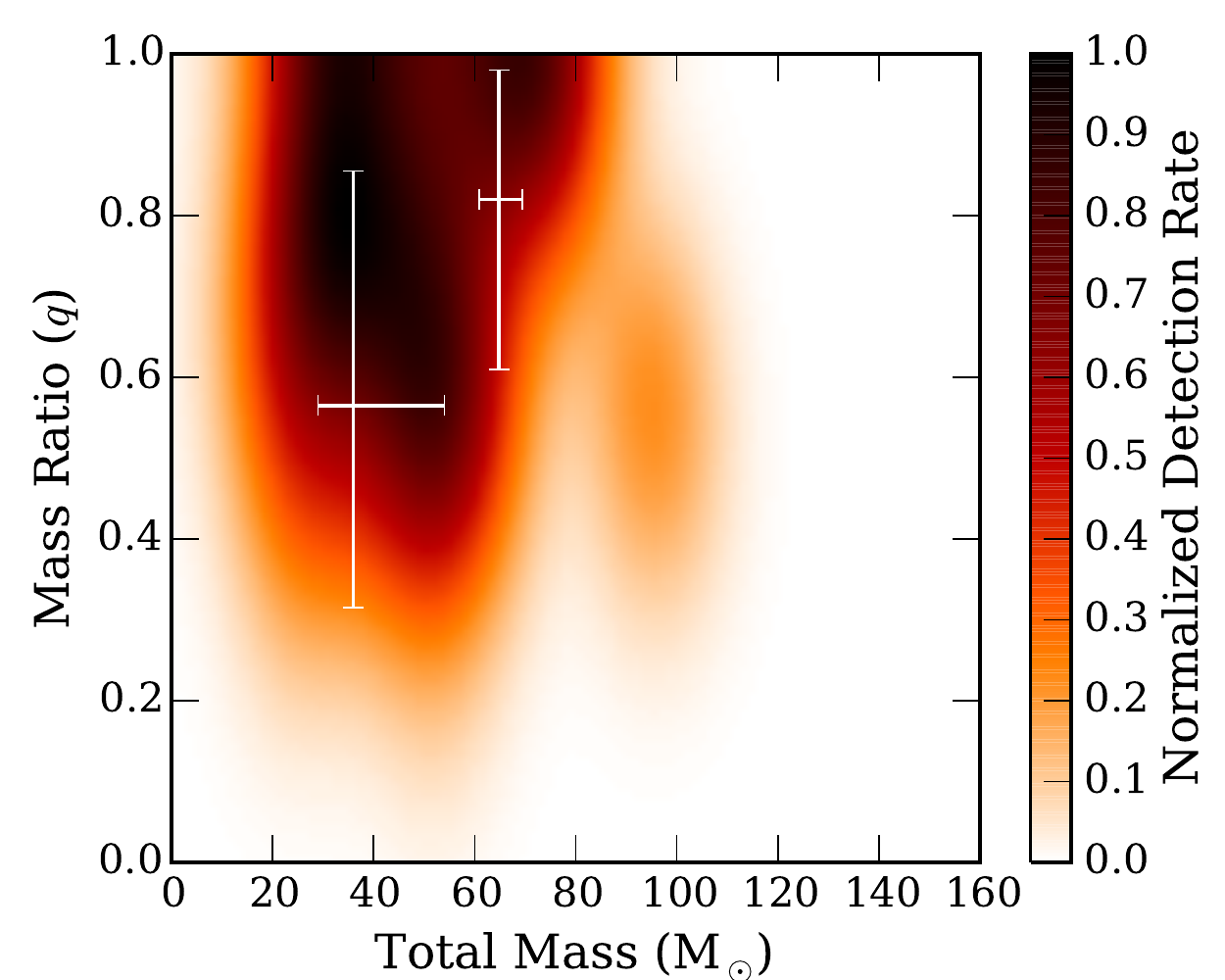}\\
\caption{\label{fig:agemassrat} Total binary mass and mass ratio distributions for inspirals for an $\alpha=2.0$ power-law mass function from an old population of clusters. {\em Left panels:} The distribution of dynamically formed BH-BH binaries in a single star cluster for IMF $M^{-2}$ that merge within the last $2.5\,$Gyr of the cluster's lifetime. 
{\em Right:} Normalized detection rate of inspiralling BH-BH binaries for all star clusters within the detectable Universe for the O1 run of LIGO Hanford.  The errorbars show the $90\%$ confidence interval for the properties of GW150914 and LVT151012.}
\end{figure*}

The total event rate per star cluster is then
\begin{equation}
\Gamma = \int_{M_{\min}}^{M_{\max}}\int_{M_{\min}}^{M_{\max}} B f(M_1)f(M_2) \,\mathrm{d}M_1\, \mathrm{d}M_2
\end{equation}
which depends on the total binary mass as
\begin{equation}
\frac{\partial\Gamma}{\partial M_{\rm tot}} \propto M_{\rm tot}^{\beta}
\int_{\max(M_{\rm tot}-M_{\max},M_{\min})}^{\min(M_{\rm tot}-M_{\min},M_{\max})} f(M_1) f(M_{\rm tot}-M_1)\,\mathrm{d}M_1
\end{equation}
This integral may be evaluated analytically for $f(M)\propto M^{-\alpha}$ using Gauss's hypergeometric function ${}_2F_1(a,b;c;z)$ as
\begin{equation}\label{eq:Gamma}
\frac{\partial\Gamma}{\partial M_{\rm tot}} \propto M_{\rm tot}^{1+\beta-2
\alpha}\left.z\; {}_2F_1\left(\frac12,\alpha;\frac32;z^2\right)\right|_{z_{\min}}^{z_{\max}}
\end{equation}
where 
\begin{align}
z_{\min} &= \max\left( 1-\frac{2M_{\max}}{M_{\rm tot}}, \frac{2M_{\min}}{M_{\rm tot }} - 1 \right)\\
z_{\max} &= \min\left( 1-\frac{2M_{\min}}{M_{\rm tot}}, \frac{2M_{\max}}{M_{\rm tot }} - 1 \right)
\end{align}
This may be expressed with elementary functions for integer and half-integer $\alpha$. Plots of Eq.~(\ref{eq:Gamma}) show that the merger rate peaks at $M_{\rm tot}\sim M_{\max}$ for $1\lesssim \alpha\lesssim 3.5$ and near $2M_{\min}$ for larger $\alpha$. Accounting for the correlation between $M_{\rm tot}$ and $q$ seen in Fig.~\ref{fig:massrat} skews $\partial \Gamma/\partial M_{\rm tot}$ to even higher $M_{\rm tot}$, discussed next. We discuss how this relationship may be used to probe the underlying mass distribution in Section \ref{sec:conc}.

Dynamical encounters not only preferably form high mass BH-BH binaries, but as $M_{\rm tot}$ approaches $2M_{\rm max}$ the binaries necessarily approach $q=1$. In Fig.~\ref{fig:massrat}, we show two-dimensional ($M_{\rm tot}$--$q$) distribution functions for BH-BH mergers from the cluster for two BH mass functions using our Monte Carlo results.  These distributions were recreated from the simulation results using a Gaussian kernel density estimator to generate a smooth distribution.   The top panel shows the results from our power-law mass function $M^{-2}$ and $M_{\rm max} = 40\,\msun$.  The bottom panel shows the results from the \citet{2004ApJ...611.1068B} mass function.  Both distributions peak near $q = 1.0$.  The peak is much more pronounced in mass functions with $\alpha <2$.  For mass functions with $\alpha \gtrsim 2$, the $q$ distribution marginalized over $M_{\rm tot}$ rises rapidly near $q=0.4$ and is flat for higher $q$ until $q=1.0$.  

Since the maximum distance to which aLIGO may detect a binary is different for different $M_{\rm tot}$ and $q$, the observed rate distribution of dynamically formed binaries will be biased relative to the merger rate, $\mathcal{R}=V_{\rm det} \Gamma$ (see Eq.~\ref{eq:V}). \citet{OKL09} has shown that for the planned aLIGO spectral noise density, the detection distance will be largest for $M_{\rm tot}$ between $60$ and $90\,\msun$ for circular binaries (see their figure 11). For the most recent observing run, O1, aLIGO was most sensitive to BH mergers with $M_{\rm tot} \approx 77\,\msun$. Further, the detection rate is biased toward equal-mass binaries as shown by Eq.~(\ref{eq:V}).

The overall impact of aLIGO's sensitivity can be seen by comparing the left and right panels of Figure~\ref{fig:massrat}.   In the right panel, we show the distribution of mergers weighted by the observable volume of the Universe respectively for different $(M_{\rm tot},q)$ for the aLIGO O1 observing run. All results have been normalized to peak at 1. By comparing our results to the panels on the left, we can see that aLIGO should expect to see high mass black hole binaries, with mass ratios peaking near 1.  However, in these models the overall distribution is mostly determined by the underlying dynamics of BHs in star clusters, rather then aLIGO's intrinsic sensitivity.  This suggests that Earth-based GW detectors  will be a useful probe of the underlying demographics of BHs.

So far, we have only compared the binary merger distribution to the BH IMF without regard for the age of the cluster. While young clusters may be an important contributor to the merger rate \citep{OLeary07}, it is not clear if they survive long enough to dominate the rate.  Most studies therefore focus on an old population of clusters, such as globulars, which are long lived.  In Figure ~\ref{fig:agemassrat}, we show the merger and detection rate distributions from binaries during the last $2.5\,$Gyr, roughly the observable horizon of aLIGO.  We find that mergers from old clusters is better fit by a broken power law. At high masses $M_{\rm tot} \gtrsim M_{\rm max}$, the merger rate is still enhanced by $\beta \gtrsim 4$ as in Figure~{fig:massto}.  However, for $M_{\rm tot} \lesssim M_{\rm max}$, $\beta \approx 0$.

\section{Summary and Conclusion}
\label{sec:conc}
In dense stellar systems, the black holes that form at the end stage of stellar evolution collect near the center of the cluster through dynamical friction.  Gravitational focusing in subsequent multibody dynamical interactions
naturally increase the interaction rate between the most massive black holes in the cluster, eventually leading them to merge. In this work we have used a suite of numerical simulations to better understand the properties of BH-BH binaries that are formed in these clusters.  We have found that dynamical interactions between the BHs strongly enhance the rate of mergers among higher mass BHs, such that the merger rate of binaries is boosted by a factor $\propto M_{\rm tot}^{\beta}$, with $\beta \gtrsim 4$, over a random selection of BH pairs from the cluster.

This relationship may be a useful tool to probe the underlying mass distribution of BHs in the aLIGO era with multiple detections. 
The merger distribution can be estimated directly from the underlying BH mass function, assuming that aLIGO can distinguish the two primary channels of producing BH-BH binaries, or that BH-BH binaries formed during dynamical interactions dominate the detection rate. Under this assumption, the aLIGO measurement statistics will give physical meaning to the underlying mass function and dynamics.

In this work, we have minimally explored the underlying reason for $\beta \gtrsim 4$.  In the dynamical interactions considered here, the process most sensitive to $M_{\rm tot}$ is the rate of binary formation through three-body encounters, which scales as $M_{\rm tot}^{5}$ \citep{2005MNRAS.358..572I}.  We have run our Monte Carlo simulations without any primordial binaries, and indeed found $\beta \approx 5$. In these simulations, the binaries that form are sampled from the entire BH population and should follow this distribution\footnote{Running our simulations without three-body binary formation but with primordial binaries had a best fit slope near $\beta \approx 3$}. Subsequent interactions between the new binary do not appear to alter the distribution.  In our $N$-body
simulations, we have found that $\beta \approx 4$, when we include a background potential.  However, we find $\beta \gg 4$ when this potential was absent. We expect that this much higher boosting rate was caused by mass segregation in the core of the BH cluster, which is not present in our Monte Carlo simulations.   Presently, our simulations do not include the impact of relaxation between the BH cluster and background stellar population. Nevertheless, the results presented in  \citet{2015ApJ...800....9M}, who used a similar initial mass function, appear consistent with our results.

Overall, we find the most likely BH binaries to be detected by aLIGO  have total masses slightly less than $\sim 2 M_{\rm max}$ when $\alpha \lesssim 2$. For shallow mass functions, with $\alpha < 2$, the effect is most prominent.  For the case where $\beta \approx 4$ and $\alpha \approx 2$, the distribution of BH mergers is fairly broad between $M_{\rm max}$ and $2M_{\rm max}$. Mass segregation in the core of the BH cluster, however, can significantly boost the fraction of mergers near $2 M_{\rm max}$.  
Note that aLIGO is sensitive to inspirals with masses up to $400\,\msun$ and it may measure the ringdown waveform for even higher $M_{\rm tot}$ \citep{2007CQGra..24S.689A,2011PhRvD..83l2005A,2015PhRvD..91l4042K}. Thus LIGO will be capable of measuring the value of $M_{\max}$ over a wide range.
A Bayesian analysis of the distribution of LIGO detections will allow one to put constraints on the underlying initial BH mass function and degree of mass segregation in these clusters. 

Looking at the distribution of mergers in Figure~\ref{fig:massrat}, there is a small population of mergers with $M_{\rm tot} > 2M_{\rm max}$.  These binaries must have formed from at least one BH that was involved in a previous merger.  For low-spinning BHs, as we consider here, we expect low merger kick velocities when $q\approx 1$, as these binaries emit GWs symmetrically \citep{2010ApJ...719.1427V}. These merger remnants then remain in the cluster as the largest BH, and subsequently merge with another high mass BH in the cluster. These mergers are therefore distributed around $(M_{\rm tot},q)=(2.5 M_{\rm max}, 0.5)$. Since BH spin is expected to reach $a=0.7\pm 0.1$ in circular mergers of nonspinning BHs, the heavy binary component in this population may be expected to have such high spin while the other component is nonspinning, provided that high stellar-mass BHs form with nearly zero spin \citep{2015arXiv151204897A}.
These subsequent mergers are infrequent, $\approx 7\,\%$ of mergers for the O1 science run of aLIGO, however they represent a smoking gun of dynamical interactions. At the final design specifications aLIGO will be more sensitive at detecting this population of BHs, $11\,\%$ of all inspirals in dense stellar clusters will constitute subsequent mergers.  For an old population of clusters, the aLIGO detection rate of subsequent mergers is closer to $15\,\%$ of all detected inspirals.

\acknowledgments{
RMO acknowledges the support provided by NSF grant AST-1313021. This work was supported in part by the European Research Council under the European Union's Horizon 2020 Programme, ERC-2014-STG grant GalNUC 638435, and it was completed in part [by BK] in the Aspen Center for Physics, which is supported by NSF grant \#PHY-1066293. The N-body simulations were carried out on the NIIF HPC cluster at the University of Debrecen, Hungary.}


\end{document}